# Deformation Behavior of multi-spherulitic nylon6/silica nanocomposites

Saeid Arabnejad, Nhan Tien Cao, VPW Shim

Department of Mechanical Engineering National University of Singapore, 9 Engineering Drive 1, Singapore 117575, Singapore

Abstract

In this study, the deformation of nylon6/silica nanocomposite is investigated by employing a multiscale computational approach to understand the influence of nanoparticles. Initially, the upper and lower bounds for the elastic properties of a combination of crystalline and amorphous lamella are predicted via Voigt and Reuss model. Subsequently, these results are used in an FEM model for RVEs representing the multi-spherulitic structure of nylon6 and a silica/nylon6 nanocomposite. Each spherulite in these models has directional mechanical properties defined by spherical coordinates. Simulation of deformation applied in orthogonal directions and the mechanical response of the pure polymer and nanocomposite are examined. The results show that spherical nanoparticles have a smaller potential for enhancement of mechanical response compared to nanoparticles of other shapes.

Keywords: semi-crystalline polymer, multiscale, nanocomposite

Introduction

The use of polymers by industry is increasing because of polymer properties such as relatively low weight and high toughness. As a result, the production of polymers has had an average annual growth rate of 8.1% [1] in recent years. Among polymers, semi-crystalline one have the highest consumption. As their name implies, this group of polymers is neither completely crystalline nor amorphous, but contains a mixture of ordered crystalline and randomly oriented amorphous regions.

The mechanical behavior of semi-crystalline polymers can be divided into elastic and inelastic responses where different mechanisms are involved in each of them. Like other semi-crystalline polymers, the deformation of nylon6 involves influences ranging from molecular-scale interactions to microstructural effect [2]. It is instructive to establish a link between deformation mechanisms at the atomic scale and macro-scale mechanical behavior. Although there has been progress towards this goal, establishment of a reliable relationship between deformation in semi-crystalline polymers at the atomic-scale and that at the macro-scale for elastic and inelastic loading is still a major challenge. Addressing all mechanisms via a single scale of simulation is not possible because of limitations in the size or details in modeling.

In semi-crystalline polymer nanocomposites, the introduction of nanoparticle into the polymer matrix alters the deformation mechanism for elastic and inelastic deformation, and therefore results in different mechanical responses. Computational techniques are useful in modeling the behavior of nanocomposites where interactions at the atomic level are altered by nanoparticles. The present study employs a hierarchical multi-scale computational approach to model the elastic mechanical response of a nylon6-silica nanocomposite to understand mechanisms involved in altering the mechanical behavior.

## Micro-structure of semi-crystalline nanocomposites

The structure of semi-crystalline nylon6 can be modeled as layered composite comprising two phases –crystalline and amorphous –in lamella form (Fig. 1e). These adjacent connected layers have different mechanical properties – the crystalline phase is strongly anisotropic, while the amorphous phase is isotropic. The crystalline phase may contain the α or γ form; however, the α phase is energetically more favorable [3]. Both α and γ phases display anisotropic elastic properties, with the stiffness in the chain direction being at least an order of magnitude higher than the perpendicular directions [4]. The amorphous phase, which consists of an assembly of randomly oriented polymer chains, shows isotropic behavior. There is strong bonding between the two layers (i.e. phases) because of the chains which extend from one phase to the other.

The properties of semi-crystalline polymers depend on a number of parameters, particularly on the degree of crystallinity and initial conformation of its spherulitic structure. The degree of crystallinity in a semi-crystalline polymer is the fraction of material which is crystalline. Based on the fabrication method and conditions, four microstructure morphologies may arise in semi-crystalline polymers – isotropic spherulitic, oriented spherulites, oriented structure and shish-kebab structure [5]. The focus of this study is on micro-structures comprising isotropic spherulites as the main contributor to the overall behavior of nanocomposite.

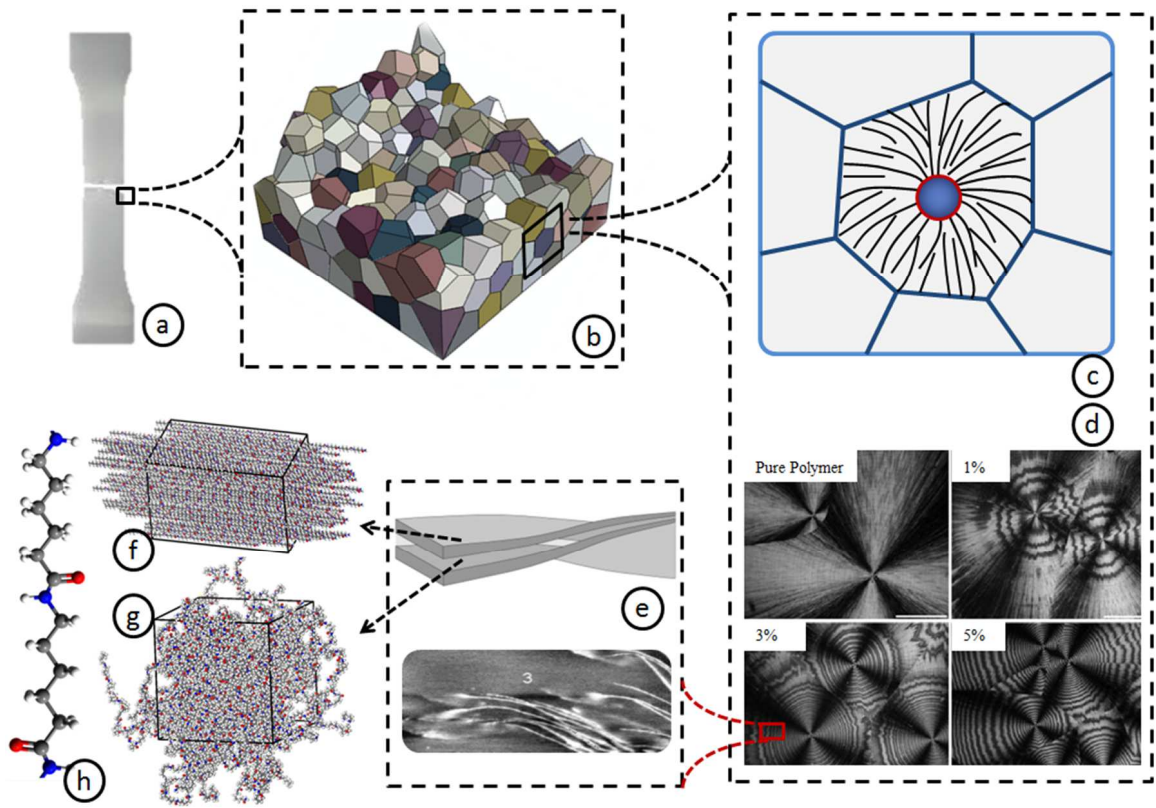

Figure 1 Structural hierarchy of semi-crystalline polymers and spherical particle reinforced nanocomposites (a) dog-bone specimen (b) tessellation of spherulites (c) location of a nanoparticle in spherulite and direction of crystalline lamella (d) change in the microstructure by addition of nanoparticles [6] (e) twist in crystalline lamella during crystallization [7] (f) crystalline phase (g) amorphous phase (h) atomic conformation of nylon6.

Crystallization in semi-crystalline polymers such as nylon6 leads to the formation of a spherulite structure, where assemblies of crystalline and amorphous layers emanate from the center of the spherulite which is, the seed point – i.e. spherulites are formed by a radial arrangement of crystalline lamella [8]. This radial arrangement of crystalline lamella with strong anisotropic properties results in non-uniform deformation within the structure, even for uniform loading [9]. Uchida et al. studied the elastic and plastic deformation of a semi-crystalline polymer using a penalty method for the deformation of crystalline lamella in the chain direction [10], [11] and concluded that the distribution of material orientation in the spherulite plays a key role in the deformation behavior. In nanocomposites with a semi-crystalline matrix and spherical nanoparticles, crystallization starts from the nanoparticles and results in a morphology with smaller spherulites (Figure 1.d) [6].

Plastic deformation in semi-crystalline polymers involves different complicated mechanisms – e.g. slip in the crystalline phase. Micromechanical modeling, considering plastic deformation and texture evolution have been developed by van Dommelen et al. [12] and Nikolov et al. [13] based on the inclusion model introduced by Lee et al. [14]. The Taylor[15] or Sachs[16] models has been used in these studies to model the mechanical properties of randomly oriented inclusions of amorphous and crystalline lamellae. Although the micromechanical models mentioned considered plastic slip in deformation, they did not include all possible deformation mechanisms possible in crystals and the multi-spherulitic structure.

# Multiscale modeling

Use of polarized optical microscopy to observe a thin layer of nylon6 melt enables observation of the process of polymer crystallization. In an experiment to define its micro-structure, nylon6 is dissolved in toluene and a thin layer of the solution is then placed on a glass lamella. This lamella is then attached to the temperature controlled hot stage of a polarized microscope (Nikon Eclipse LV100). By increasing the temperature of the stage, the toluene evaporates and the remaining nylon6 melts. This thin layer of nylon6 melt is then cooled at a rate of 1˚C/min to crystallize. This process results in the multi-spherulitic structure shown in Figure 2. In each spherulite, brighter and darker radial lines of the cells represent the amorphous and crystalline region respectively. When a beam of light is passed through the polymer film, the crystalline phase diffracts the beam and results in a darker line, while the amorphous phase does not change the beam direction, resulting in brighter lines in the image.

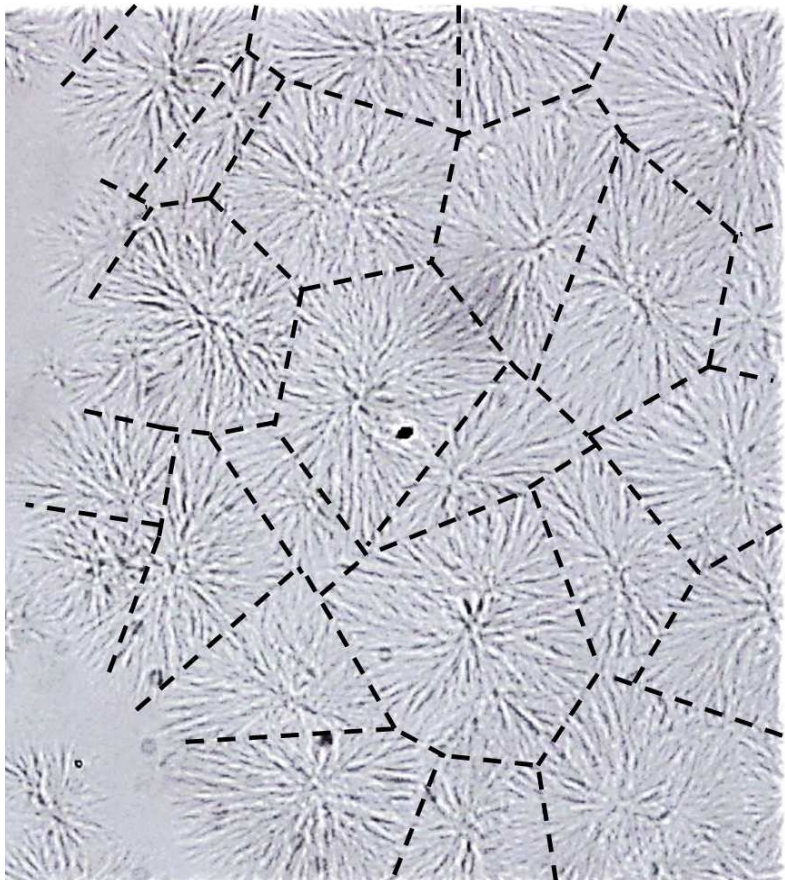

Figure 2 Microstructure of a thin layer of nylon6 captured using a hot-stage polarized microscope (dashed lines indicate the boundary of the spherulites).

The generation of models begin by assigning the seed points randomly in representative vomule elements (RVEs). It is noted that for nanocomposite models, the seed points – i.e. the locations of nanoparticles – are sufficiently far apart for nanoparticles not to overlap. When the locations of seed points in the model are defined, Voronoi cells representing polymer spherulites are generated using the Voro++ package [17]. The geometries of the Voronoi cells – i.e. the coordinates of planes forming the cell – are then processed by a Matlab script to generate a Python script, which is then incorporated into the Abaqus finite element package [20], to generate cell tessellations. Figure 3 shows four randomly generated conformations for multi-spherulitic nylon6, containing 25 spherulites, while

Figure 4 shows four models with randomly generated conformations for nylon6/silica nanocomposite containing 25 spherulites with nanoparticles embedded within them.

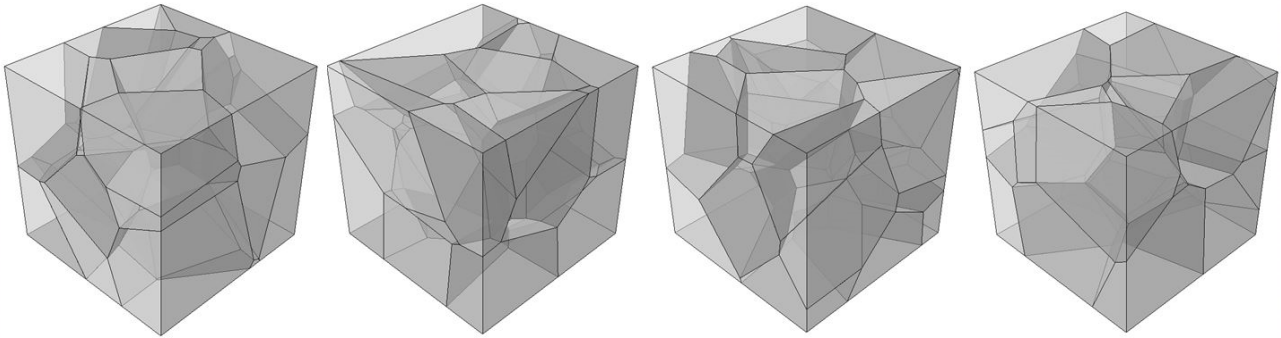

Figure 3 RVE of four random configurations of pure nylon6 comprising 25 spherulites.

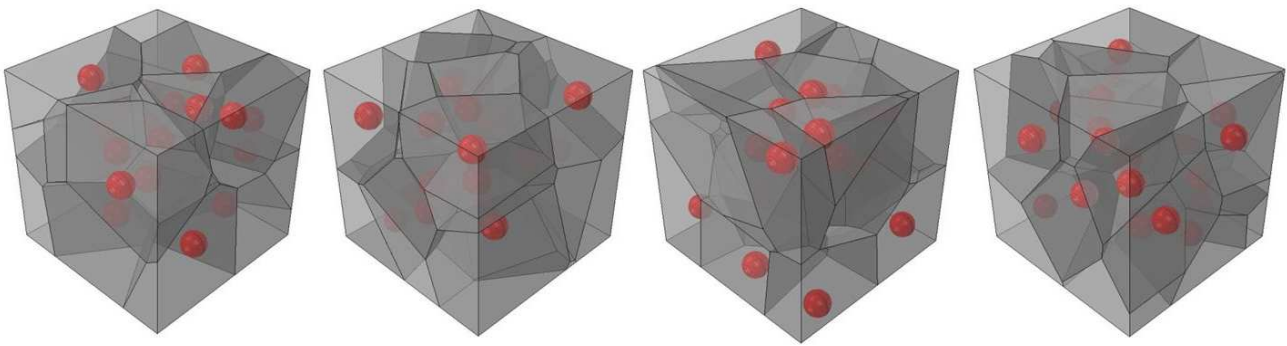

Figure 4 RVE of four random configurations of nanocomposites comprising 25 spherulites.

To define directional properties for the polymer matrix, the origin of a local spherical coordinate system for each of the spherulites is defined at its center (Figure 5). As shown in Figure 6, the stiffness in direction 1 (chain direction) is different from that in the 2 and 3 directions. In order to simplify the effect of twist in crystalline lamella, equal values for the stiffness in directions 2 and 3 are considered for the combination of amorphous and crystal lamellae. The Voronoi cells and their corresponding particles are then tied together by constraint equations, and the cells are tied, since polymer chains extend from one cell to another. The model is then discretized using tetrahedral elements, and boundary conditions corresponding to simple tension and compression are applied. An implicit analysis procedure is then employed to elicit the response of these models.

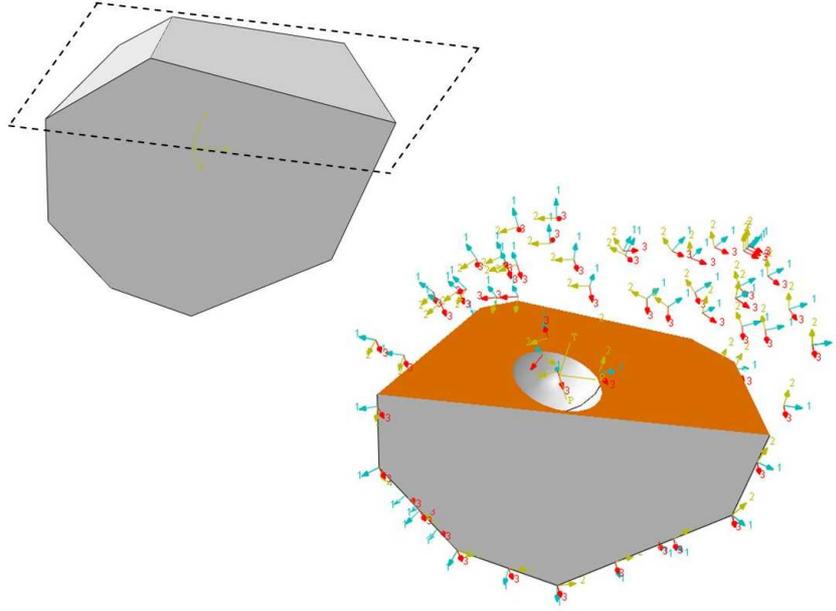

Figure 5 Spherical coordinate system defined for each cell to assign directional material properties.

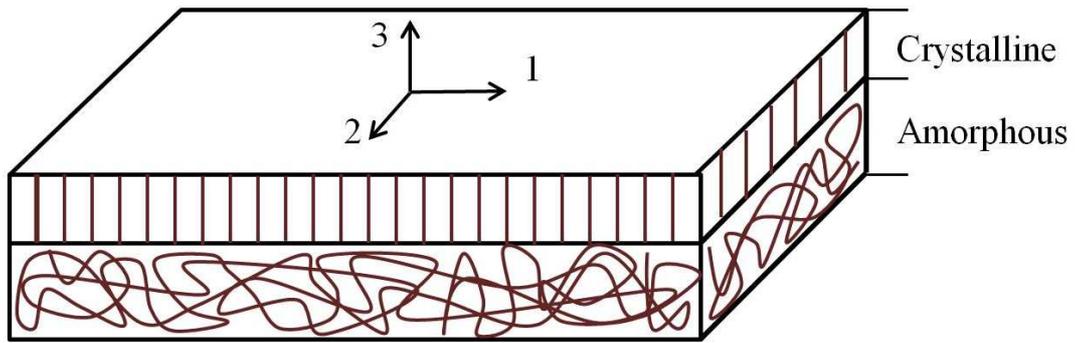

Figure 6 Schematic diagram of lamellar conformation of the crystalline and amorphous phases of the polymer.

The material properties of the polymer matrix are obtained using the rule of mixture, based on the properties of their component phases. The Voigt and Reuss models [18] are employed to determine the overall behavior of the combination, and these correspond respectively to upper and lower bound approaches. Assuming affine deformation conditions for the different phases of bilayer model in Fig. 6, the Voigt method defines the overall elastic stiffness tensor $\underline{C}$, of a combination of amorphous and crystalline phase as:

$$\underline{C} = (1-f)\underline{C}^a + f\,\underline{C}^c, \qquad (1)$$

where f is the degree of crystallinity, and $\underline{C}^a$ and $\underline{C}^c$ are the stiffness tensors of the amorphous and crystalline phases respectively. The degree of crystallinity in this study is considered to be 40% [19] which is the volume fraction of the crystalline phase. If the stress in each phase is the same, the Reuss approach expresses the overall stiffness tensor of the inclusion, $\underline{C}$, as follows:

$$\frac{1}{\underline{C}} = (1-f)\frac{1}{\underline{C}^a} + f\,\frac{1}{\underline{C}^c} \qquad (2)$$

The actual state of a composite is actually neither uniform in stress nor strain with respect to its constituents; thus, the stiffness of a composite has a value between these two bounds.

Table 1 Directional elastic stifnesses of bilayer amorphous-crystalline model, based on Voigt and Reuss models.

|  | $C_{22}$ and $C_{33}$ (GPa) | $C_{11}$ (GPa) |
|---|---|---|
| Voigt | 80.2 | 20.2 |
| Reuss | 4.83 | 4.79 |

Based on DFT_D calculations, the elastic constants computed for crystalline nylon in tension, are $E_{11} = 45$ GPa, $E_{22} = 27.9$ GPa, and $E_{33} = 367$ GPa; while for compression $E_{11} = 47$ GPa, $E_{22} = 37$ GPa, and $E_{33} = 352$ GPa [20]. Since a complicated state of stress occurs in each spherulite and the elastic properties in tension and compression are quite similar, the values for the elastic modulus in tension and compression are averaged to the represent the elastic behavior of the crystalline phase. Subsequently, the elastic constants of the bilayer model are calculated using the Voigt and Reuss methods, and these are presented in Table 1. The stiffness in the radial direction in each spherulite is defined as the mixture of stiffness of amorphous phase and crystalline lamella in hydrogen bond direction. Because of twist in lamellae in a spherulite (Fig. 1d), the stiffness in the other two directions are considered as an average of amorphous-crystal bilayer model stiffnesses in directions 3 and 2 – 196 GPa.

## Results and Discussion

Each set of results presented for stress-strain response is the average of four random Voronoi tessellations.

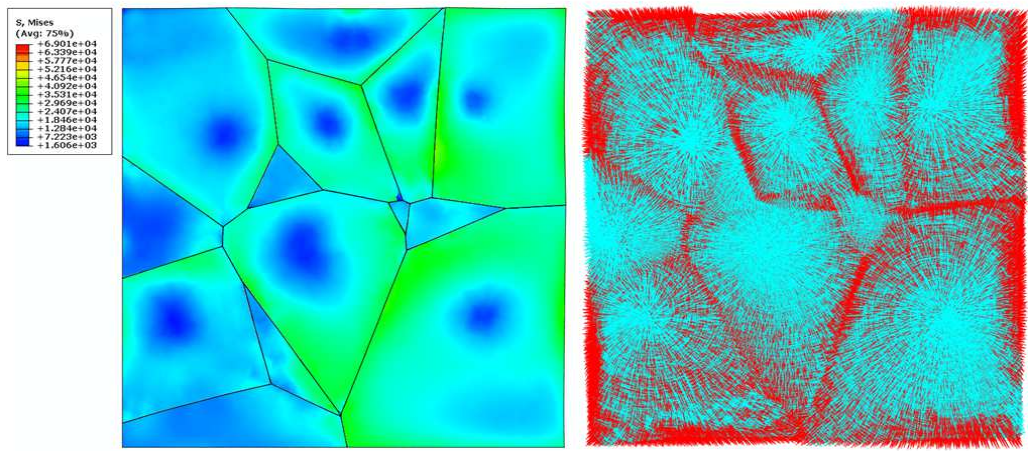

Figure 7(left) von Mises stress distribution in a multi-spherulitic structure of pure nylon6 under compression and (right) material orientation for the same section (red lines show crystallization direction – direction 1 in Figure 6).

The stress distribution in a cross-section of multi-spherulitic nylon6 is depicted in Figure 7. The center of the spherulites experiences smaller values of stress compared to the boundary regions which shows that the central portion contributes less to load transfer during deformation. As crystallization semi-crystalline polymers in nanocomposites initiate from the nanoparticles, these particles are located at the center of spherulites [6]. Therefore, introducing a nanoparticle into the central region of a spherulite, whereby a smaller contribution to load transfer is carried out, may have a small influence on stiffness enhancement. As shown in the experimentally measured response of silica/nylon6 nanocomposites [21], addition of silica nanoparticles does not enhance the mechanical behavior of nylon6 significantly.

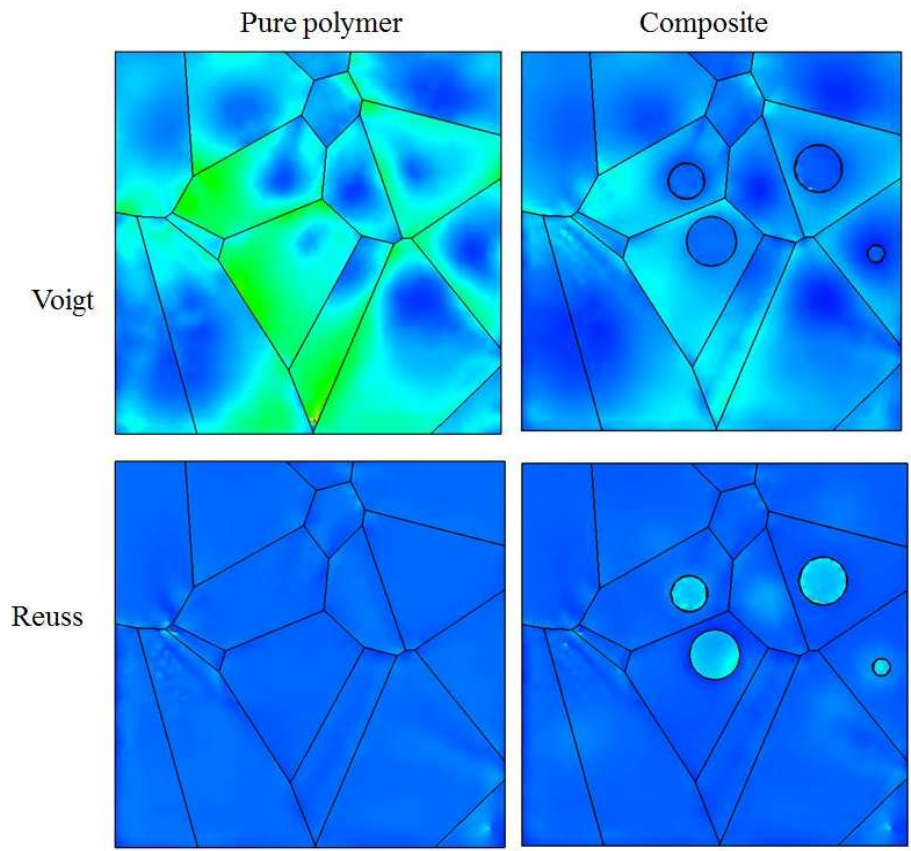

Figure 8 von Mises stress distribution in pure nylon6 and nanocomposite under compression.

Figure 8 shows the stress distribution in the cross-section of a polymer and nanocomposite model predicted by the Voigt and Reuss approaches. The pattern of stress distribution suggests that when a polymer is anisotropic, the center of each cell in pure polymer and the nanoparticles in a composite do not experience high stress. However, a semi-crystalline polymer corresponding to the Reuss model behaves the same as an isotropic amorphous model, with a uniform stress distribution, and the nanoparticles in the composite experience higher stresses than that of the matrix.

Figure 9 shows the upper and lower bond for the mechanical response of multi-spherulitic nylon6 and its nanocomposite in tension, while Figure 10 shows responses for compression. All models simulate deformation up to 10% engineering strain.

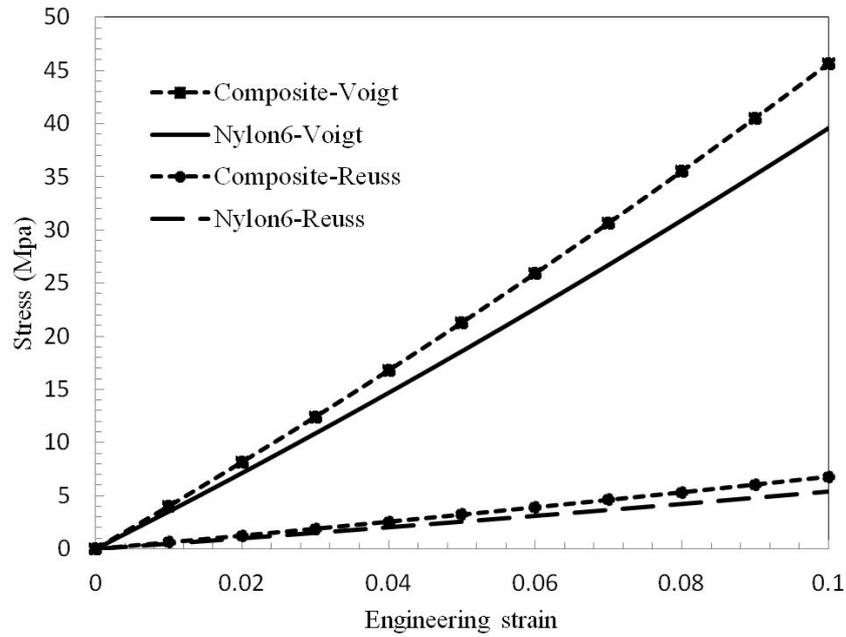
Figure 9 Stress-strain behavior of pure polymer and composite under tension.

As depicted in Fig. 9 and 10, simulations predict that the introduction of silica nanoparticles into nylon6 enhances mechanical properties of polymer in compression and tension; however, this enhancement is relatively larger for the models adopting the Reuss rule of mixtures. The simulation results do not show any significant geometric non-linearity in the mechanical response for elastic deformation.

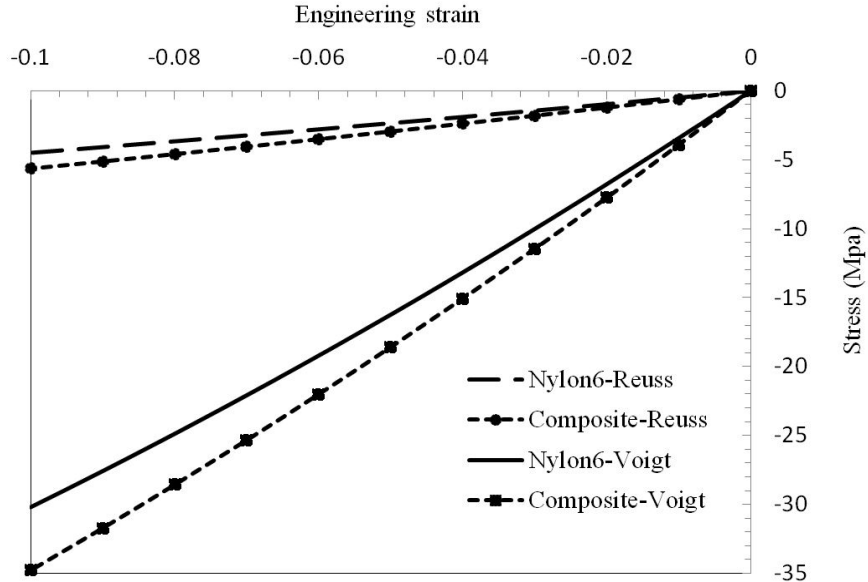
Figure 10 Stress-strain behavior of pure polymer and nanocomposite models for compression.

## Conclusions

The mechanical behavior of multi-spherulitic nylon6 and nylon6/silica nanocomposite is analyzed using a multiscale modeling technique. Simulation results show that when anisotropic behavior is considered for the bilayer model comprising crystalline and amorphous lamella via the Voigt model, nylon6 experiences a smaller stress at the core of each spherulite, and most of the load is transferred via the outer layers of spherulites. However, adopting Reuss model, the results show a more uniform stress distribution. Modelling a nanocomposite using a Voigt approach – the upper bound for mechanical behavior – shows a smaller amount of stress in nanoparticles compared to the outer layers of spherulites and suggests that the nanoparticles in these nanocomposites are not incorporated in an adequate location to help in load transfer. The modeling technique presented in this article, can be used for other semi-crystalline polymers and nanocomposites with spherical inclusions.

# References

[1] F. PARDOS, "Plastics in the twenties," *ANTEC … Conf. Proc.*, vol. 3, pp. 3638–3641.

[2] Z. Bartczak and a. Galeski, "Plasticity of Semicrystalline Polymers," *Macromol. Symp.*, vol. 294, no. 1, pp. 67–90, Aug. 2010.

[3] S. Dasgupta, W. B. Hammond, and W. Goddard, "Crystal structures and properties of nylon polymers from theory," *J. Am. Chem. Soc.*, vol. 118, pp. 12291–12301, 1996.

[4] Y. Li and W. Goddard, "Nylon 6 crystal structures, folds, and lamellae from theory," *Macromolecules*, vol. 35, no. 22, pp. 8440–8455, 2002.

[5] L. Lin and A. S. Argon, "Deformation resistance in oriented nylon 6," *Macromolecules*, vol. 25, no. 15, pp. 4011–4024, Jul. 1992.

[6] C. Yao and G. Yang, "Poly(trimethylene terephthalate)/silica nanocomposites prepared by dual in situ polymerization: synthesis, morphology, crystallization behavior and mechanical properties," *Polym. Int.*, vol. 59, no. 4, pp. 492–500, Apr. 2010.

[7] J. Xu, B.-H. Guo, Z.-M. Zhang, J.-J. Zhou, Y. Jiang, S. Yan, L. Li, Q. Wu, G.-Q. Chen, and J. M. Schultz, "Direct AFM Observation of Crystal Twisting and Organization in Banded Spherulites of Chiral Poly ( 3-hydroxybutyrate-co-3-hydroxyhexanoate )," *Macromolecules*, vol. 37, no. 11, pp. 4118–4123, 2004.

[8] L. Lin and A. S. Argon, "Structure and plastic deformation of polyethylene," *J. Mater. Sci.*, vol. 29, no. 2, pp. 294–323, 1994.

[9] M. F. Butler and A. M. Donald, "Deformation of spherulitic polyethylene thin films," *J. Mater. Sci.*, vol. 32, no. 14, pp. 3675–3685, 1997.

[10] M. Uchida, T. Tokuda, and N. Tada, "Finite element simulation of deformation behavior of semi-crystalline polymers with multi-spherulitic mesostructure," *Int. J. Mech. Sci.*, vol. 52, no. 2, pp. 158–167, Feb. 2010.

[11] Y. Tomita and M. Uchida, "Computational characterization of micro- to mesoscopic deformation behavior of semicrystalline polymers," *Int. J. Mech. Sci.*, vol. 47, no. 4–5, pp. 687–700, Apr. 2005.

[12] J. a. W. van Dommelen, D. M. Parks, M. C. Boyce, W. a. M. Brekelmans, and F. P. T. Baaijens, "Micromechanical modeling of intraspherulitic deformation of semicrystalline polymers," *Polymer (Guildf).*, vol. 44, no. 19, pp. 6089–6101, Sep. 2003.

[13] S. Nikolov, R. Lebensohn, and D. Raabe, "Self-consistent modeling of large plastic deformation, texture and morphology evolution in semi-crystalline polymers," *J. Mech. Phys. Solids*, vol. 54, no. 7, pp. 1350–1375, Jul. 2006.

[14] B. Lee, D. M. Parks, and S. Ahzi, "Micromechanical modeling of large plastic deformation and texture

evolution in semi-crystalline polymers," *J. Mech. Phys. Solids*, vol. 41, pp. 1651–1687, 1993.

[15] G. I. Taylor, "Plastic strain in metals," *J. Inst. Met.*, vol. 62, pp. 307–324, 1938.

[16] G. Sachs, "Zur ableitung einer Fliessbedingung," *Zeitschrift des Vereins Dtsch. Ingenieure fur Maschinenbau und Met.*, vol. 72, pp. 734–736, 1928.

[17] C. H. Rycroft, "VORO++: a three-dimensional voronoi cell library in C++.," *Chaos*, vol. 19, no. 4, p. 041111, Dec. 2009.

[18] J. Qu and M. Cherkaoui, *Fundamentals of Micromechanics of Solids*. New Jersey: Wiley, 2006.

[19] T. J. Bessell, D. Hull, and J. B. Shortall, "Effect of Polymerization Conditions and Crystallinity on Mechanical-Properties and Fracture of Spherulitic Nylon-6," *J. Mater. Sci.*, vol. 10, no. 7, pp. 1127–1136, 1975.

[20] S. Arabnejad, S. Manzhos, C. He, and V. Shim, "Shear-induced conformation change in α-crystalline nylon6," *Appl. Phys. Lett.*, vol. 105, no. 22, p. 221910, Dec. 2014.

[21] H. Gu, Y. Guo, S. Y. Wong, C. He, X. Li, and V. P. W. Shim, "Effect of interphase and strain-rate on the tensile properties of polyamide 6 reinforced with functionalized silica nanoparticles," *Compos. Sci. Technol.*, vol. 75, pp. 62–69, Feb. 2013.